\documentclass[english,aps,preprint]{revtex4}
\usepackage[T1]{fontenc}
\usepackage[latin9]{inputenc}
\usepackage{amsmath}
\usepackage{graphicx}
\usepackage{amssymb}

\providecommand{\tabularnewline}{\\}

\usepackage{babel}

\begin{document}

\title{RESONANT PRODUCTION OF COLOR OCTET ELECTRON AT THE LHeC}

\author{M. SAHIN}

\email{m.sahin@etu.edu.tr}

\affiliation{TOBB University of Economics and Technology, Physics Division, Ankara,
Turkey}

\author{S. SULTANSOY}

\email{ssultansoy@etu.edu.tr}

\affiliation{TOBB University of Economics and Technology, Physics Division, Ankara,
Turkey}

\affiliation{Institute of Physics, National Academy of Sciences, Baku, Azerbaijan}

\author{S. TURKOZ}

\email{turkoz@science.ankara.edu.tr}

\affiliation{Ankara University, Department of Physics, Ankara, Turkey}
\begin{abstract}
In composite models with colored preons leptogluons ($l{}_{8}$) has
a same status with leptoquarks, excited leptons and quarks etc. We
analyze resonant production of color octet electron ($e_{8}$) at
QCD Explorer stage of the Large Hadron electron Collider (LHeC). It
is shown that the $e_{8}$ discovery at the LHeC simultaneously will
determine the compositeness scale. 
\end{abstract}
\maketitle

\section{INTRODUCTION}

A large number of {}``fundamental'' particles, as well as observable
free parameters (put by hand), in Standard Model (SM) indicate that
it is not {}``the end of story''. Physics has met similar situation
two times in the past: one is the Periodic Table of the Elements which
was clarified by Rutherford's experiment later, the other is hadron
inflation which has resulted in quark model. This analogy implies
the preonic structure of the SM fermions (see \cite{I.A. D'SOUZA}
and references therein). The preonic models predict a zoo of new particles
such as excited leptons and quarks, leptoquarks, leptogluons etc.
Excited fermions and leptoquarks are widely discussed in literature
and their searches are inseparable parts of future collider's physics
programs. Unfortunately, leptogluons did not attract necessary attention,
while they are predicted in all models with colored preons (see, for
example, \cite{HModel}, \cite{FM Model}, \cite{GS Model}, \cite{BM Model},
\cite{BS Model}, \cite{CKS Model}).

Lower bound on leptogluon masses, $86$ GeV, given in PDG \cite{PDG}
reflects twenty years old Tevatron results \cite{Abe}. As mentions
in \cite{Hewett} D0 clearly exclude $200$ GeV leptogluons and could
naively place the constraint $M_{LG}\gtrsim325$ GeV. Fifteen years
old H1 results on color octet electron, $e{}_{8}$, search \cite{H1}
has excluded the compositeness scale $\Lambda\lesssim$ $3$TeV for
$M_{e8}\backsimeq100$ GeV and $\Lambda\lesssim$$240$ GeV for $M_{e8}\backsimeq250$
GeV. The advantage of lepton-hadron colliders is the resonant production
of leptogluons, whereas at hadron and lepton colliders they are produced
in pairs.

The sole realistic way to TeV scale in lepton-hadron collisions are
presented by linac-ring type electron-proton colliders (see reviews
\cite{EPS 2003}, \cite{PAC 2005}, \cite{Akay} and references therein).
Recently CERN, ECFA and NuPECC initiated the study on the LHC based
ep colliders \cite{LHeC web}. Two options are considered for the
Large Hadron electron Collider (LHeC): the construction of new e-ring
in the LHC tunnel \cite{Dainton} or the construction of e-linac tangentially
to the LHC \cite{Zimmerman}, \cite{Epac 2006}, \cite{Epac 2008}.
It should be noted that energy of electrons in first option is limited
by synchrotron radiation, whereas in second option energy of electrons
can be increased by lengthening the linac. Tentative parameters for
linac-ring options of the LHeC are presented in the Table 1. QCD Explorer
stage(s) is mandatory: it will provide necessary information on PDF's
for adequate interpretation of future LHC results and it will clarify
QCD basics, as well. The realization of the Energy Frontier stage(s)
will be determined by the LHC results on Beyond the Standard Model
(BSM) physics.

\begin{table}
\begin{tabular}{|c|c|c|c|}
\hline 
Stage  & $E_{e},$GeV  & $\sqrt{s},$TeV  & $L$, $10^{32}$$cm^{-2}s^{-1}$\tabularnewline
\hline
\hline 
LHeC/QCDE-1  & $70$  & $1.4$  & $1-10$\tabularnewline
\hline 
LHeC/QCDE-2  & $140$  & $1.98$  & $1-10$\tabularnewline
\hline 
LHeC/EF  & $500$  & $3.74$  & $1$\tabularnewline
\hline
\end{tabular}

\caption{Tentative parameters of the LHeC linac-ring options. QCDE and EF denotes
QCD Explorer and Energy Frontier, respectively.}

\end{table}

In this paper we investigate potential of QCDE stages of the LHeC
in search for color octet electron via resonant production. In section
2, Lagrangian for $e{}_{8}$ interactions is presented and it's decay
widths and production cross sections at different stages of LHeC are
evaluated. Section 3 is devoted to detailed analysis of leptogluon
signatures at QCD-E stages of the LHeC. Finally, concluding remarks
are given in section 4.

\section{INTERACTION LAGRANGIAN, DECAY WIDTH AND PRODUCTION CROSS SECTION}

For the interaction of leptogluons with corresponding lepton and gluon
we use the following Lagrangian \cite{PDG}, \cite{Kantar 1998}:

\begin{equation}
L=\frac{1}{2\Lambda}\underset{l}{\sum}\left\{ \bar{l}_{8}^{\alpha}g_{s}G_{\mu\nu}^{\alpha}\sigma^{\mu\nu}(\eta_{L}l_{L}+\eta_{R}l_{R})+h.c.\right\} \end{equation}

where $G_{\mu\nu}^{\alpha}$ is field strength tensor for gluon, index
$\alpha=1,2,...,8$ denotes the color, $g_{s}$ is gauge coupling,
$\eta_{L}$ and $\eta_{R}$ are the chirality factors, $l_{L}$ and
$l_{R}$ denote left and right spinor components of lepton, $\sigma^{\mu\nu}$
is the anti-symmetric tensor and $\Lambda$ is the compositeness scale.
The leptonic chiral invariance implies $\eta{}_{L}$$\eta_{R}=0$.
For numerical calculations we add leptogluons into the CalcHEP program
\cite{CalcHEP}.

Decay width of the color octet electron is given by

\begin{equation}
\Gamma_{e8}=\frac{\alpha_{s}M_{e8}^{3}}{4\Lambda^{2}}\end{equation}

In Fig. 1 decay widths of leptogluons are presented for two scenarios,
$\Lambda=M_{e8}$ and $\Lambda=10$ TeV. %
\begin{figure}
\includegraphics[scale=0.7]{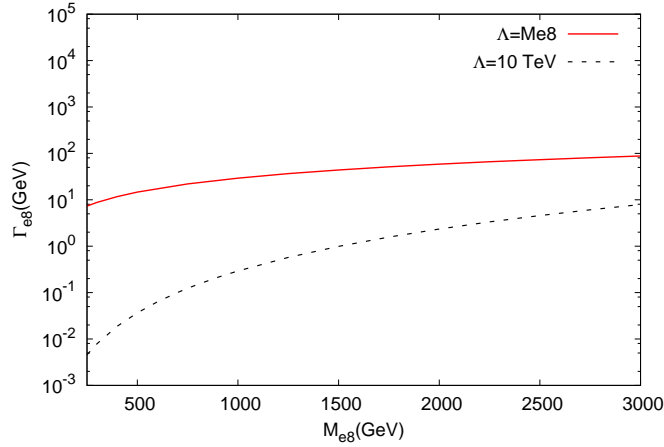}

\caption{Leptogluon decay width via its mass for $\Lambda=M_{e8}$ and $\Lambda=10$
TeV. }

\end{figure}

The resonant $e_{8}$ production cross sections for there stages at
the LHeC from Table 1, evaluated using CalcHEP with CTEQ6L parametrization
\cite{CTEQ6L} for parton distribution functions, are presented in
Figs. 2-4. It is seen that sufficiently high cross-sections allow
the exploration of the $e_{8}$ mass range almost up to the kinematical
limits.

\begin{figure}
\includegraphics[scale=0.7]{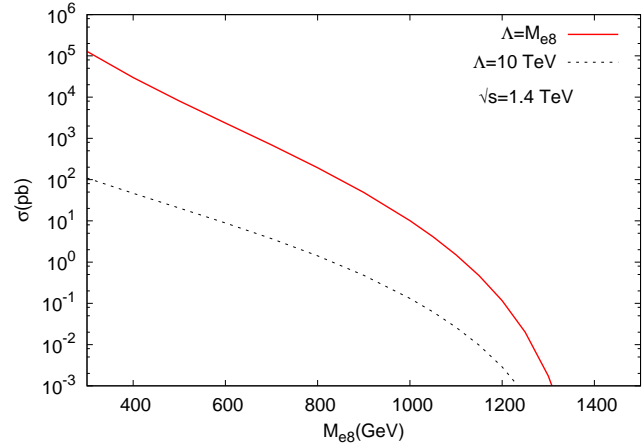}

\caption{Resonant $e_{8}$ production at the LHeC/QCDE-1.}

\end{figure}

\vspace{10cm}

\begin{figure}
\includegraphics[scale=0.7]{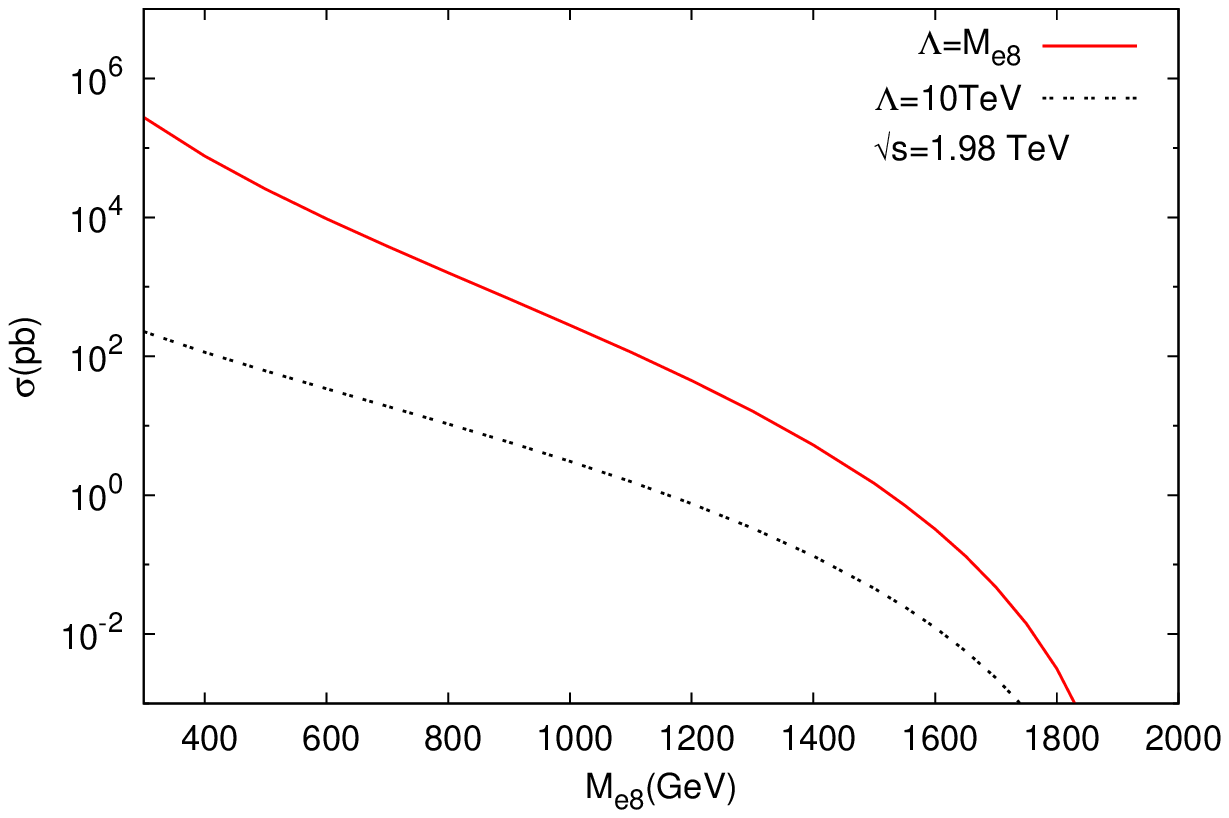}

\caption{Resonant $e_{8}$ production at the LHeC/QCDE-2.}

\end{figure}

\vspace{-3cm}

\begin{figure}
\includegraphics[scale=0.7]{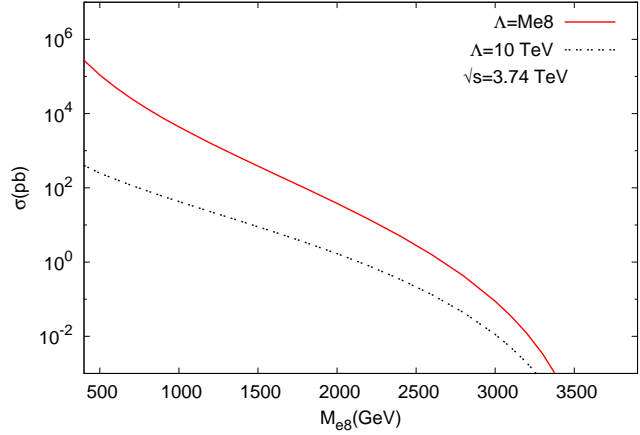}

\caption{Resonant $e_{8}$ production at the LHeC/EF.}

\end{figure}

\vspace{-3cm}

\section*{III. SIGNAL AND BACKGROUND ANALYSIS}

\subsection{LHeC/QCDE-1 stage}

First of all, let us consider $p_{T}$ and $\eta$ distributions for
signal and background processes in order to determine appropriate
kinematical cuts. Transverse momentum distribution of final state
jets for signal at $\Lambda=10$ TeV and background is shown in Fig.
5. It is seen that $p_{T}>150$ GeV cut essentially reduces background,
whereas signal is almost unaffected. Figs. 6 and 7 represent pseudo-rapidity
($\eta$) distributions for electrons and jets, respectively. As seen
from figure 7, $\eta_{e^{-}}$distribution for signal and background
are not drastically different. Concerning $\eta_{j}$, most of signal
lies above $\eta=0$, whereas 99 \% of background is concentrated
in $-2<\eta_{j}<0$ region. For this reasons below we use $p_{T}>150$,
$|\eta_{e}|<4$ and $0<\eta_{j}<4$. With these cuts we present in
Fig. 8 invariant mass distributions for signal and background.

\begin{figure}
\includegraphics[scale=0.7]{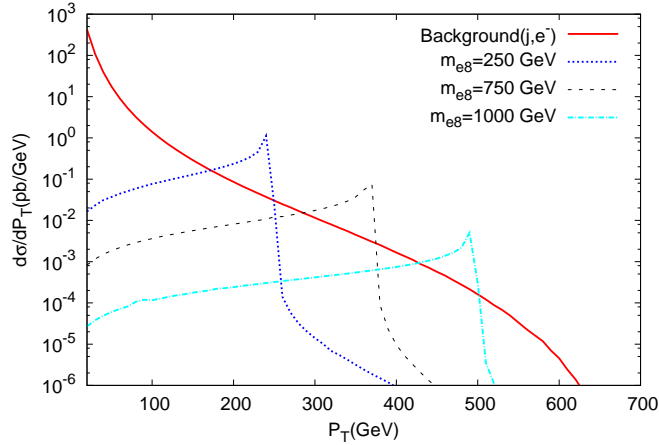}

\caption{Transverse momentum distributions of final state jets for signal and
background at $\sqrt{s}=1.4$ TeV and $\Lambda=10$ TeV. }

\end{figure}

\begin{figure}
\includegraphics[scale=0.7]{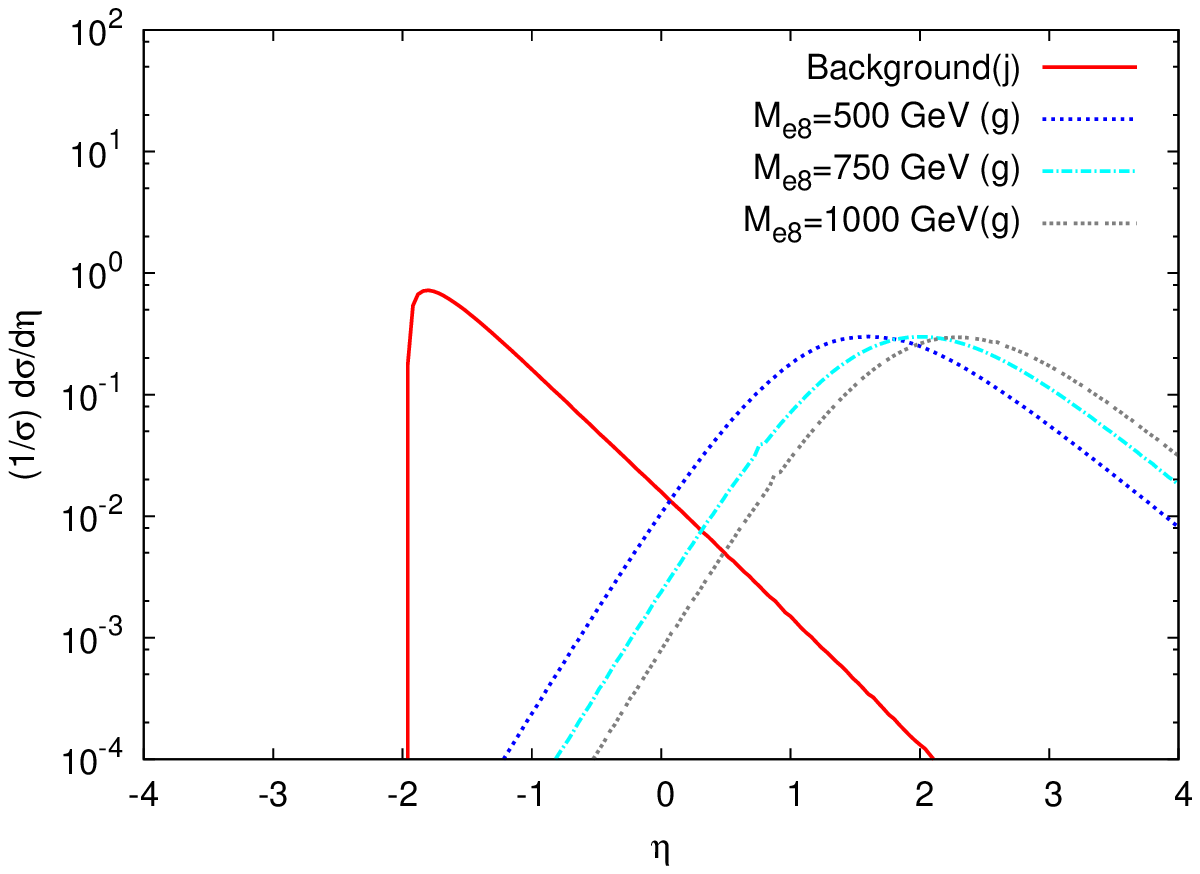}

\caption{Pseudo-rapidity distributions of jets for signal and background at
$\sqrt{s}=1.4$ TeV and $\Lambda=10$ TeV. }

\end{figure}

\vspace{10cm}

\begin{figure}
\includegraphics[scale=0.7]{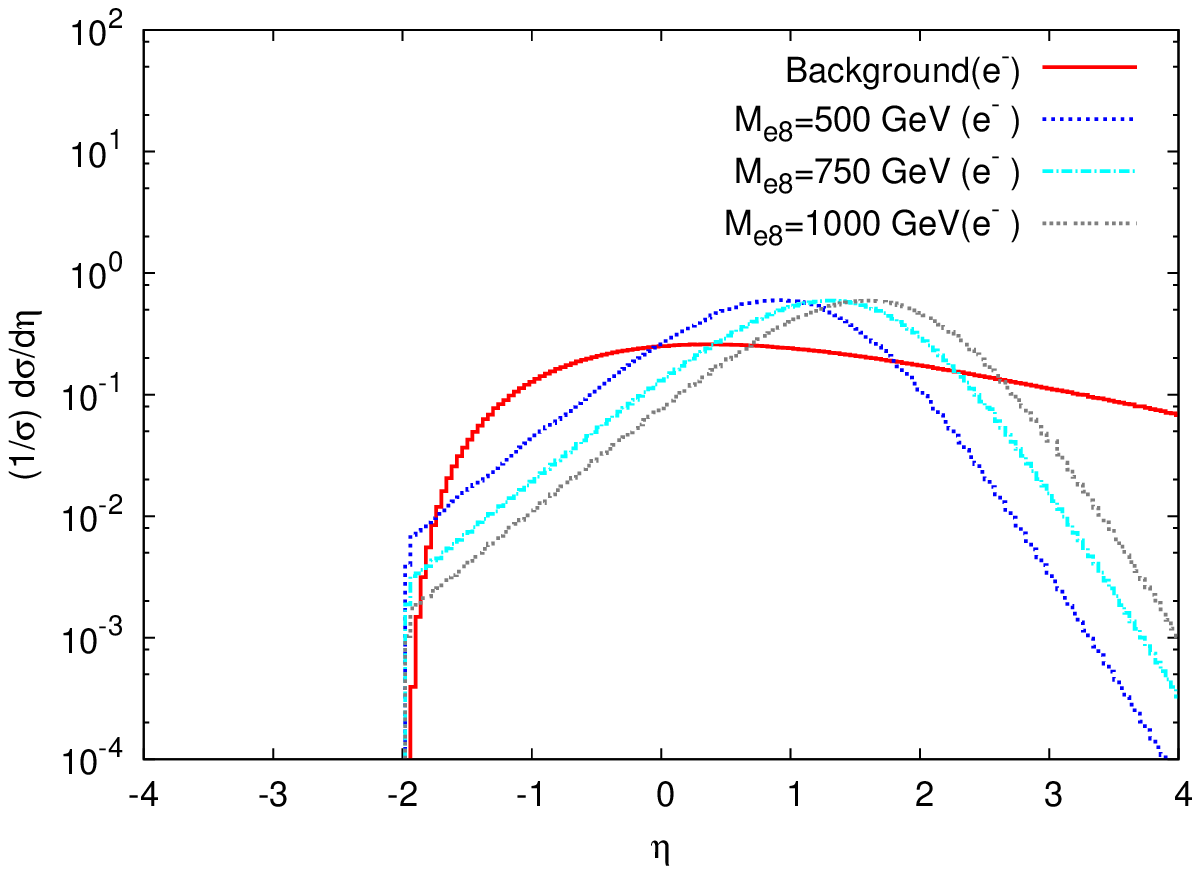}

\caption{Pseudo-rapidity distributions of electrons for signal and background
at $\sqrt{s}=1.4$ TeV and $\Lambda=10$ TeV.}

\end{figure}

\vspace{-5cm}

\begin{figure}
\includegraphics[scale=0.7]{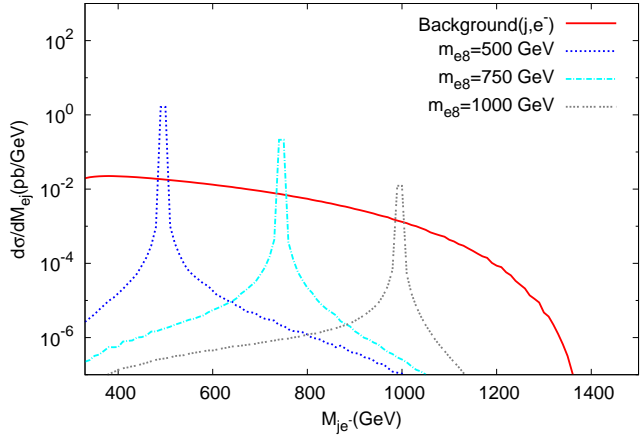}

\caption{$ej$ invariant mass distributions for signal and background at $\sqrt{s}=1.4$
TeV and $\Lambda=10$ TeV. }

\end{figure}

\vspace{-5cm}

Advantage of resonant production will provide an opportunity to probe
compositeness scale will above the center of mass energy of the collider.
For statistical significance we use following formula:

\begin{equation}
SS=\frac{S}{\sqrt{S+B}}\end{equation}

where $S$ and $B$ denote number of signal and background events,
respectively.

Numbers of signal and background events for different $M_{e8}$ values
are presented in Table 2 for $L_{int}=1$$fb^{-1}$. In calculating
these values, in addition to cuts given above, we have used mass windows
as $M_{e8}-2\Gamma_{e8}<M_{ej}<M_{e8}+2\Gamma_{e8}$ for $\Gamma_{e8}>10$
GeV and $M{}_{e8}$ $-20$ GeV $<M_{ej}<M_{e8}+20$ GeV for $\Gamma{}_{e8}<10$
GeV. It is seen that resonant production of color octet electron will
provide very clean signature for masses up to $M_{e8}\simeq1$TeV.

In Table 3 we present reachable compositeness scale values for $L_{int}=1$
and $L_{int}=10$$fb^{-1}$. It is seen that multi-hundred TeV scale
can be searched for $M_{e8}=500$ GeV. Then, increase of the luminosity
by one order results in two times higher values for $\Lambda$.

Lastly, for a given $L_{int}=1$$fb^{-1}$, the upper mass limits
for $5\sigma$ discovery at LHeC/QCDE-1 stage are $M_{e8}=1100$ GeV
and $M_{e8}=1275$ GeV for $\Lambda=10$ TeV and $\Lambda=M_{e8}$,
respectively.

\begin{table}
\begin{tabular}{|c|c|c|c|c|}
\hline 
$M_{e8},$ GeV  & \multicolumn{2}{c|}{$\wedge=M_{e8}$} & \multicolumn{2}{c|}{$\wedge=10$ TeV}\tabularnewline
\hline
\hline 
 & $S$  & $B$  & $S$  & $B$\tabularnewline
\hline 
$500$  & $1.1\times10^{7}$  & $1.1\times10^{3}$  & $3.3\times10^{4}$  & $700$\tabularnewline
\hline 
$750$  & $6.4\times10^{5}$  & $630$  & $4.2\times10^{3}$  & $280$\tabularnewline
\hline 
$1000$  & $2.2\times10^{4}$  & $\mbox{1}65$  & $250$  & $53$\tabularnewline
\hline 
$1250$  & $81$  & $6$  & $1$  & $1$\tabularnewline
\hline
\end{tabular}

\caption{Number of signal and background events for LHeC/QCDE-1 with $L_{int}=1$$fb^{-1}$. }

\end{table}

\begin{table}
\begin{tabular}{|c|c|c|}
\hline 
$M_{e8},$ GeV  & $L_{int}=1$$fb^{-1}$  & $L_{int}=10$$fb^{-1}$\tabularnewline
\hline
\hline 
$500$  & $150$ ($200$)  & $275$ ($350$)\tabularnewline
\hline 
$750$  & $65$ ($90$)  & $125$ ($160$)\tabularnewline
\hline 
$1000$  & $22$ ($30$)  & $45$ ($58$)\tabularnewline
\hline
\end{tabular}

\caption{Achievable compositeness scale ($\Lambda$ in TeV units) at LHeC/QCDE-1
for $5\sigma$ ($3\sigma$) statistical significance.}

\end{table}

\subsection{LHeC/QCDE-2 stage}

In order to determine corresponding cuts we present $p_{T}$, $\eta_{j}$
and $\eta_{e}$ distributions for signal and background processes
in Figs 9, 10 and 11, respectively. The figures indicate that significant
change takes place only for $\eta_{j}$. In this subsections we will
use $p_{T}>150$, $|\eta_{e}|<4$ and $-0.5<\eta_{j}<4$. The invariant
mass distributions obtained with this cuts are given in Fig. 12.

\begin{figure}
\includegraphics[scale=0.7]{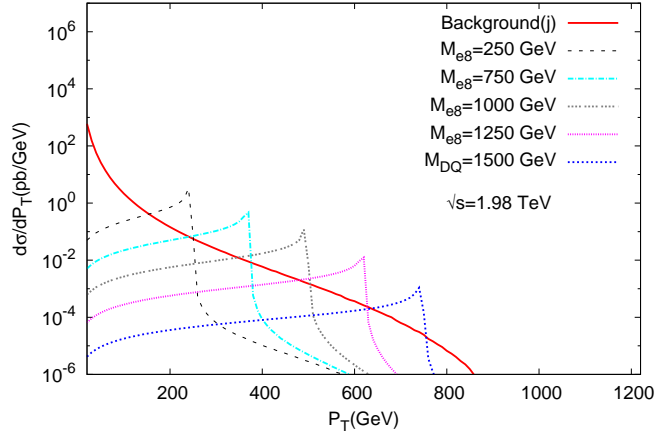}

\caption{Transverse momentum distributions of final state jets for signal and
background at $\sqrt{s}=1.98$ TeV and $\Lambda=10$ TeV. }

\end{figure}

\begin{figure}
\includegraphics[scale=0.7]{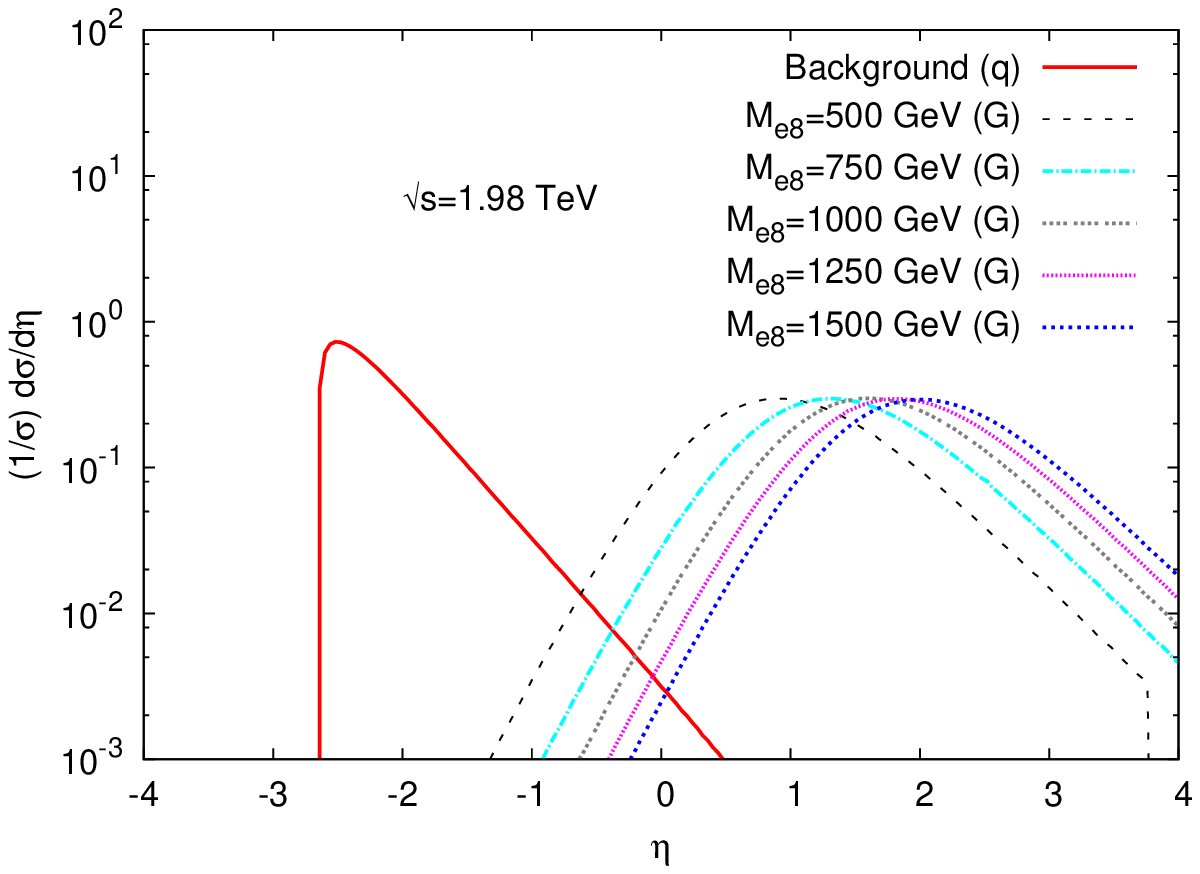}

\caption{Pseudo-rapidity distributions of jets for signal and background at
$\sqrt{s}=1.98$ TeV and $\Lambda=10$ TeV. }

\end{figure}


%
\begin{figure}
\includegraphics[scale=0.7]{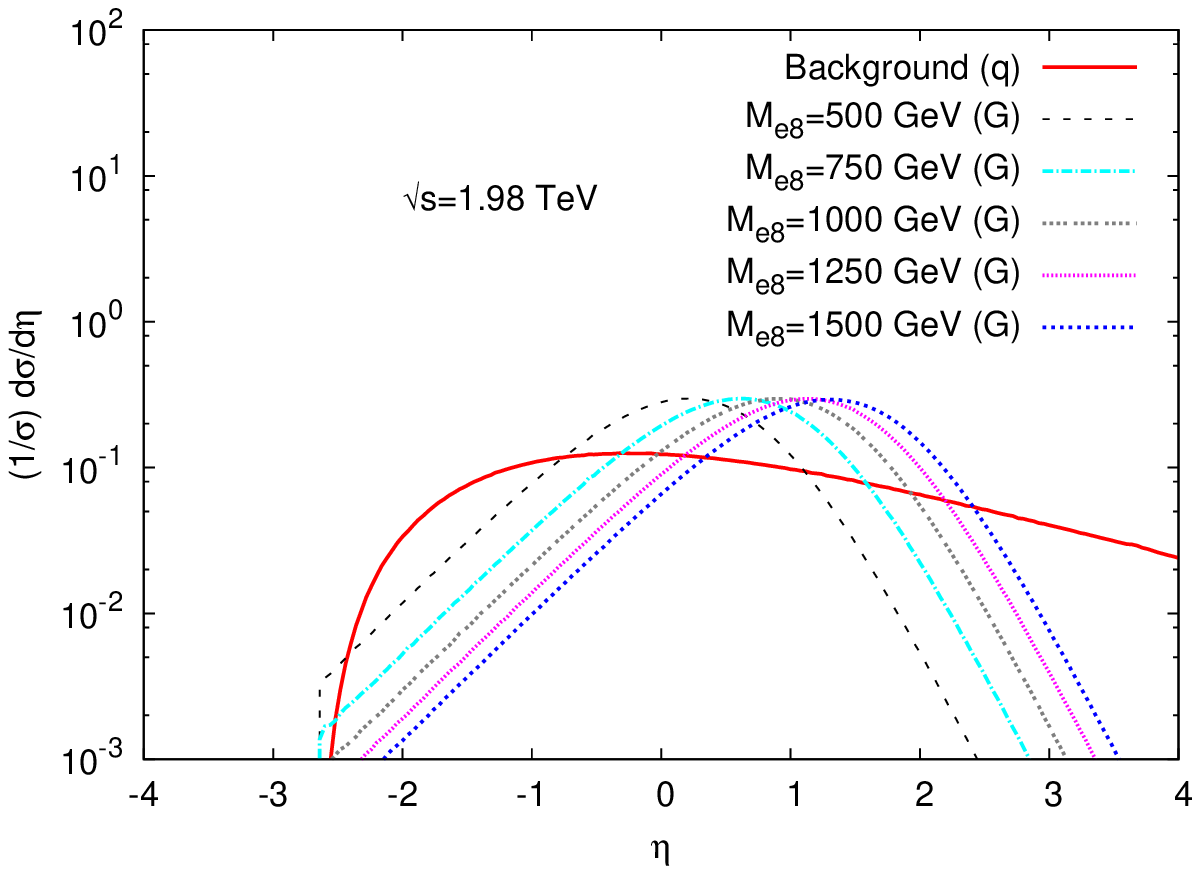}

\caption{Pseudo-rapidity distributions of electrons for signal and background
at $\sqrt{s}=1.98$ TeV and $\Lambda=10$ TeV.}

\end{figure}

\begin{figure}
\includegraphics[scale=0.7]{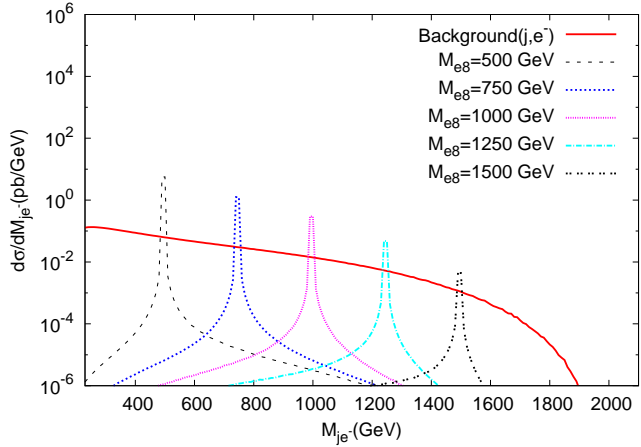}

\caption{$ej$ invariant mass distributions for signal and background at $\sqrt{s}=1.98$
TeV and $\Lambda=10$ TeV. }

\end{figure}

The numbers of signal and background events for 6 different $M_{e8}$
values are presented in Table 4 (the mass window used is the same
as the one used in previous subsection). As seen from the Table very
clean signal can be obtained up to $M_{e8}\backsimeq1500$ GeV.

Reachable $\Lambda$ scales for $5$ different mass values are given
Table 5. Comparison with Table 3 show that twofold increasing of the
electron energy results in: $1.5$ times higher values of $\Lambda$
for $M_{e8}=500$ GeV, $2$ times - for $M_{e8}=750$ GeV and $4$
times - for $M_{e8}=1000$ GeV. Moreover, multi-ten TeV scales can
be achieved for $M_{e8}=1250$ and $1500$ GeV, which are not available
at LHeC/QCDE-1. 

Finally, for a given $L_{int}=1$$fb^{-1}$, the upper mass limits
for $5\sigma$ discovery at LHeC/QCDE-2 are $M_{e8}=1580$ GeV and
$M_{e8}=1775$ GeV for $\Lambda=10$ TeV and $\Lambda=M_{e8}$, respectively.

\vspace{10cm}

\begin{table}
\begin{tabular}{|c|c|c|c|c|}
\hline 
$M_{e8},$ GeV  & \multicolumn{2}{c|}{$\wedge=M_{e8}$} & \multicolumn{2}{c|}{$\wedge=10$ TeV}\tabularnewline
\hline
\hline 
 & $S$  & $B$  & $S$  & $B$\tabularnewline
\hline 
$500$  & $3.3\times10^{7}$  & $1.4\times10^{3}$  & $9.8\times10^{4}$  & $940$\tabularnewline
\hline 
$750$  & $3.9\times10^{6}$  & $1000$  & $2.6\times10^{4}$  & $445$\tabularnewline
\hline 
$1000$  & $5.0\times10^{5}$  & $630$  & $5.8\times10^{3}$  & $210$\tabularnewline
\hline 
$1250$  & $5.3\times10^{4}$  & $270$  & $980$  & $76$\tabularnewline
\hline 
$1500$  & $3.5\times10^{3}$  & $77$  & $89$  & $16$\tabularnewline
\hline 
$1750$  & $55$  & $6$  & $2$  & $1$\tabularnewline
\hline
\end{tabular}

\caption{Number of signal and background events for LHeC/QCDE-2 with $L_{int}=1$$fb^{-1}.$}

\end{table}

\vspace{-5cm}


%
\begin{table}
\begin{tabular}{|c|c|c|}
\hline 
$M_{e8},$ GeV  & $L_{int}=1$$fb^{-1}$  & $L_{int}=10$$fb^{-1}$\tabularnewline
\hline
\hline 
$500$  & $245$ ($320$)  & $440$ ($570$)\tabularnewline
\hline 
$750$  & $150$ ($195$)  & $275$ ($355$)\tabularnewline
\hline 
$1000$  & $82$ ($110$)  & $155$ ($205$)\tabularnewline
\hline 
$1250$  & $41$ ($56$)  & $81$ ($107$)\tabularnewline
\hline 
$1500$  & $16$ ($23$)  & $34$ ($46$)\tabularnewline
\hline
\end{tabular}

\caption{Achievable compositeness scale ($\Lambda$ in TeV units) at LHeC/QCDE-2
for $5\sigma$ ($3\sigma$) statistical significance.}

\end{table}



\section*{IV. CONCLUSION}

It seems that QCD Explorer stage(s) of the LHeC, together with providing
necessary information on PDF's and QCD basics, could play essential
role on the BSM physics, also. Concerning color octet electrons. LHeC/QCDE-1
will cover $M_{e8}$ mass up to O($1200$ GeV), whereas LHeC/QCDE-2
will enlarge covered mass range up to O($1700$ GeV).

The discovery of $e_{8}$ at this machine, simultaneously will determine
compositeness scale. For example, if $M_{e8}=500$ GeV LHeC/QCDE-2
with $L_{int}=10$ $fb^{-1}$ will be sensitive to $\wedge$ up to
$570$ TeV.
\begin{acknowledgments}
Authors are grateful to A. Celikel and M. Kantar for useful discussions.
This work is supported by TUBITAK in the framework of the BIDEP post-doctoral
program and TAEK under the grant No CERN-A5.H2.P1.01-11. \end{acknowledgments}

\end{document}